\documentclass[10pt]{article}

\usepackage{epsf}
\usepackage{amsmath}
\usepackage{amsfonts}
\usepackage{amssymb}
\usepackage{bbm}
\usepackage{theorem}
\usepackage{multicol}
\usepackage{color}
\usepackage{graphicx}
\usepackage{pstricks,pst-3dplot}
\usepackage{pst-plot}
\usepackage{pst-node}
\usepackage{pst-grad}
\usepackage{pst-coil}
\usepackage{multido}
\usepackage{mathrsfs}
\usepackage[all]{xy}
\usepackage[footnotesize,ruled]{caption}
\usepackage[hang,nooneline]{subfigure}

\newpsobject{showgrid}{psgrid}{subgriddiv=1,griddots=10,gridlabels=6pt}

\definecolor{dark-green}{rgb}{0,0.7,0}
\definecolor{dark-blue}{rgb}{0,0.2,0.5}
\definecolor{med-blue}{rgb}{0,0.7,1}
\definecolor{mblue}{rgb}{0,0.2,1}
\definecolor{cnc}{rgb}{0.8,0,0}
\definecolor{light-red}{rgb}{1,0.8,0.8}
\definecolor{dark-yellow}{rgb}{1,0.8,0}
\definecolor{light-blue}{rgb}{0.8,0.9,1}
\definecolor{verylight-blue}{rgb}{0.93,0.95,1}
\definecolor{light-yellow}{rgb}{1,0.9,0.8}

\begin{document}

\title{Charged particle interferometry in Pleba\'{n}ski--Demia\'{n}ski black hole space--times}

\author{Valeria Kagramanova$^1$, Jutta Kunz$^1$ and Claus L\"ammerzahl$^2$ \\
\\
$^1$
Institut f\"ur Physik, Universit\"at Oldenburg,
26111 Oldenburg, Germany\\
$^2$
ZARM, Universit\"at Bremen, Am Fallturm,
28359 Bremen, Germany \vspace{0.5cm}\\
\footnotesize{E-mail: kavageo@theorie.physik.uni-oldenburg.de, kunz@theorie.physik.uni-oldenburg.de,}\\ \footnotesize{laemmerzahl@zarm.uni-bremen.de}}

\maketitle

\begin{abstract}
The Pleba\'{n}ski--Demia\'{n}ski solution is a very general axially symmetric analytical solution of the Einstein field equations generalizing the Kerr solution. This solution depends on seven parameters which under certain circumstances are related to mass, rotation, cosmological constant, NUT parameter, electric and magnetic charges, and acceleration. In this paper we present a general description of matter wave interferometry in the general Pleba\'{n}ski--Demia\'{n}ski black hole space--time. Particular emphasis is placed on a gauge invariant description of the symmetries of the gauge field. We show that it is possible to have access to all parameters separately except the acceleration. For neutral particles there is only access to a combination of electric and magnetic charge.  
\end{abstract}



\section{Introduction}

A given space--time can be explored by dynamical systems like massive point particles and light rays and, indeed, most of the interpretation of the properties of a given space--time or a particular solution of Einstein's field equation is based on that \cite{Ehlers06}. A further system which may be used is quantum matter. With quantum matter not only the trajectory of the particle is subject to observation but the amplitude and the phase of the quantum systems evolving in the given space-time. In its quasiclassical limit the phase is proportional to the proper time along the classical trajectory and, thus, contains all the information about the space--time metric. This phase can be accessed through interferometry. Since the phase is very sensitive to external fields like the gravitational field, interferometry is a very precise tool to explore the properties of space--times.

The first interference experiment sensitive to the gravitational field has been carried through with neutrons in 1975 by Colella, Overhauser and Werner \cite{COW75}. They used a macroscopic interferometer: after coherent splitting neutrons moved along different paths with a height difference of several cm. With neutron interferometry also the rotation of the Earth has been observed \cite{Werneretal79}. This is a matter wave analogue of the famous Sagnac effect. Both effects have also been observed with much higher accuracy using atomic beam interferometry \cite{PetersChungChu99,GustavsonLandraginKasevich00}.

In all these cases it is sufficient to calculate the phase shift on a non--relativistic level, that is, by using the Schr\"odinger equation. However, as the accuracy of these devices increases and also interferometry with ultracold atoms with much higher accuracy in space is feasible it might be interesting to study these interference experiments in a broader context. This broader context might be (i) to include relativistic effects and effects due to the spin of particles \cite{AudretschLaemmerzahl83,BordeHouardKarasievich01,Borde01}, (ii) to include terms violating the Einstein Equivalence Principle \cite{Laemmerzahl98}, (iii) to include a general space--time metric in the frame of a post--Newtonian parametrized test theory \cite{Laemmerzahl96a}, or (iv) to use a particular space--time metric given by a certain solution of Einstein's equations. In this paper we calculate the phase shift for a charged scalar field in a Pleba\'{n}ski--Demia\'{n}ski space--time.

The Pleba\'{n}ski--Demia\'{n}ski family of solutions of the Einstein field equations  associated with the gravitational fields of isolated massive objects \cite{PlebanskiDemianski76} are known to completely exhaust the Petrov type D space--times. These axially symmetric solutions are characterized by seven parameters which under certain circumstances are related to mass, angular momentum, cosmological constant, electric and magnetic charges, NUT parameter and acceleration. This includes black hole space--times like Kerr-NUT-(A)dS space--times, the C-metric describing accelerating sources and non-expanding solutions of Kundt's class. See \cite{GriffithsPodolsky06} for a review and a new form of this family of solutions. The non--accelerating Pleba\'{n}ski--Demia\'{n}ski solutions possess the outstanding property that they allow separable Hamilton--Jacobi equations and, thus, integrability of the geodesic equation \cite{Carter68,Carter68a,DemianskiFrancaviglia80}. This also extends to higher dimensions \cite{Pageetal07,FrolovKrtousKubiznak07,KubiznakKrtous07}. Higher dimensional solutions of this type became popular in the connection of string theories and brane world models \cite{ChenLuePope06}. Here we describe charged particle interferometry in such general space--time models in order to answer the question whether all parameters characterizing the Pleba\'{n}ski--Demia\'{n}ski family of solutions are accessible through such a type of experiment.  

A scalar field obeying the Klein--Gordon equation is the simplest quantum object. Interference with quantum fields with inner degrees of freedom like spin--$\frac{1}{2}$ fields explore the same properties but are more complicated to describe. Due to their spin--curvature coupling spin--$\frac{1}{2}$ fields are more important for the local exploration of the space--time curvature \cite{AudretschLaemmerzahl83a}. However, since in ordinary situations within the Solar system the space--time curvature and, thus, the spin--curvature coupling is very small, these inner degrees of freedom effectively do not contribute to the observed phase shift. Therefore, for standard situations interferometry with scalar fields is by far sufficient. In more general geometries spin--$\frac{1}{2}$ fields play a distinctive role in the search for non--metrical fields like torsion \cite{AudretschLaemmerzahl83}. This is not considered here.

\begin{figure}[t]
\begin{center}
{\footnotesize\begin{pspicture}(-7,-3)(5,10)
\pstThreeDLine{->}(6,-4,0)(6,-4,11)
\pstThreeDLine{->}(6,-4,0)(6,0,0)
\pstThreeDLine{->}(6,-4,0)(0,-4,0)
\pstThreeDPut(-0.3,-4,0){$x$}
\pstThreeDPut(6,0.3,0){$y$}
\pstThreeDPut(6,-4,11.3){$t$}
\pstPlanePut[plane=xy,planecorr=normal](0,0,0){\rput{0}(0,0){
\psframe[linestyle=none,fillstyle=solid,fillcolor=white](-2,-1)(2,1)
\rput(-3,-0.7){source}
\rput(-1.7,-1.5){\txt{beam \\ splitter}}
\rput(2.6,-1){mirror}
\rput(-2.7,1){mirror}
\pscircle[fillstyle=solid,fillcolor=red](-3,-1){0.1}
\psline[linewidth=2pt,linecolor=gray](-2.4,-1.4)(-1.6,-0.6)
\psline[linewidth=2pt,linecolor=gray](1.6,0.6)(2.4,1.4)
\psline[linewidth=2pt,linecolor=black](-2.4,0.6)(-1.6,1.4)
\psline[linewidth=2pt,linecolor=black](1.6,-1.4)(2.4,-0.6)
\psline[linecolor=red]{->>}(-3,-1)(-2,-1)
\psline[linecolor=red]{->}(-2,-1)(2,-1)
\psline[linecolor=red]{->}(2,-1)(2,1)
\psline[linecolor=red]{->}(-2,-1)(-2,1)
\psline[linecolor=red]{->}(-2,1)(2,1)
\psline[linecolor=red]{->>}(2,1)(3,1)}}
\pstPlanePut[plane=xy,planecorr=normal](0,0,1){\rput{10}(0,0){
\psframe[linestyle=none,fillstyle=solid,fillcolor=white](-2,-1)(2,1)
\pscircle[fillstyle=solid,fillcolor=red](-3,-1){0.1}
\psline[linewidth=2pt,linecolor=gray](-2.4,-1.4)(-1.6,-0.6)
\psline[linewidth=2pt,linecolor=gray](1.6,0.6)(2.4,1.4)
\psline[linewidth=2pt,linecolor=black](-2.4,0.6)(-1.6,1.4)
\psline[linewidth=2pt,linecolor=black](1.6,-1.4)(2.4,-0.6)
\psline[linecolor=red]{->>}(-3,-1)(-2,-1)
\psline[linecolor=red]{->}(-2,-1)(2,-1)
\psline[linecolor=red]{->}(2,-1)(2,1)
\psline[linecolor=red]{->}(-2,-1)(-2,1)
\psline[linecolor=red]{->}(-2,1)(2,1)
\psline[linecolor=red]{->>}(2,1)(3,1)}}
\pstPlanePut[plane=xy,planecorr=normal](0,0,2){\rput{20}(0,0){
\psframe[linestyle=none,fillstyle=solid,fillcolor=white](-2,-1)(2,1)
\pscircle[fillstyle=solid,fillcolor=red](-3,-1){0.1}
\psline[linewidth=2pt,linecolor=gray](-2.4,-1.4)(-1.6,-0.6)
\psline[linewidth=2pt,linecolor=gray](1.6,0.6)(2.4,1.4)
\psline[linewidth=2pt,linecolor=black](-2.4,0.6)(-1.6,1.4)
\psline[linewidth=2pt,linecolor=black](1.6,-1.4)(2.4,-0.6)
\psline[linecolor=red]{->>}(-3,-1)(-2,-1)
\psline[linecolor=red]{->}(-2,-1)(2,-1)
\psline[linecolor=red]{->}(2,-1)(2,1)
\psline[linecolor=red]{->}(-2,-1)(-2,1)
\psline[linecolor=red]{->}(-2,1)(2,1)
\psline[linecolor=red]{->>}(2,1)(3,1)}}
\pstPlanePut[plane=xy,planecorr=normal](0,0,3){\rput{30}(0,0){
\psframe[linestyle=none,fillstyle=solid,fillcolor=white](-2,-1)(2,1)
\pscircle[fillstyle=solid,fillcolor=red](-3,-1){0.1}
\psline[linewidth=2pt,linecolor=gray](-2.4,-1.4)(-1.6,-0.6)
\psline[linewidth=2pt,linecolor=gray](1.6,0.6)(2.4,1.4)
\psline[linewidth=2pt,linecolor=black](-2.4,0.6)(-1.6,1.4)
\psline[linewidth=2pt,linecolor=black](1.6,-1.4)(2.4,-0.6)
\psline[linecolor=red]{->>}(-3,-1)(-2,-1)
\psline[linecolor=red]{->}(-2,-1)(2,-1)
\psline[linecolor=red]{->}(2,-1)(2,1)
\psline[linecolor=red]{->}(-2,-1)(-2,1)
\psline[linecolor=red]{->}(-2,1)(2,1)
\psline[linecolor=red]{->>}(2,1)(3,1)}}
\pstPlanePut[plane=xy,planecorr=normal](0,0,4){\rput{40}(0,0){
\psframe[linestyle=none,fillstyle=solid,fillcolor=white](-2,-1)(2,1)
\pscircle[fillstyle=solid,fillcolor=red](-3,-1){0.1}
\psline[linewidth=2pt,linecolor=gray](-2.4,-1.4)(-1.6,-0.6)
\psline[linewidth=2pt,linecolor=gray](1.6,0.6)(2.4,1.4)
\psline[linewidth=2pt,linecolor=black](-2.4,0.6)(-1.6,1.4)
\psline[linewidth=2pt,linecolor=black](1.6,-1.4)(2.4,-0.6)
\psline[linecolor=red]{->>}(-3,-1)(-2,-1)
\psline[linecolor=red]{->}(-2,-1)(2,-1)
\psline[linecolor=red]{->}(2,-1)(2,1)
\psline[linecolor=red]{->}(-2,-1)(-2,1)
\psline[linecolor=red]{->}(-2,1)(2,1)
\psline[linecolor=red]{->>}(2,1)(3,1)}}
\pstPlanePut[plane=xy,planecorr=normal](0,0,5){\rput{50}(0,0){
\psframe[linestyle=none,fillstyle=solid,fillcolor=white](-2,-1)(2,1)
\pscircle[fillstyle=solid,fillcolor=red](-3,-1){0.1}
\psline[linewidth=2pt,linecolor=gray](-2.4,-1.4)(-1.6,-0.6)
\psline[linewidth=2pt,linecolor=gray](1.6,0.6)(2.4,1.4)
\psline[linewidth=2pt,linecolor=black](-2.4,0.6)(-1.6,1.4)
\psline[linewidth=2pt,linecolor=black](1.6,-1.4)(2.4,-0.6)
\psline[linecolor=red]{->>}(-3,-1)(-2,-1)
\psline[linecolor=red]{->}(-2,-1)(2,-1)
\psline[linecolor=red]{->}(2,-1)(2,1)
\psline[linecolor=red]{->}(-2,-1)(-2,1)
\psline[linecolor=red]{->}(-2,1)(2,1)
\psline[linecolor=red]{->>}(2,1)(3,1)}}
\pstPlanePut[plane=xy,planecorr=normal](0,0,6){\rput{60}(0,0){
\psframe[linestyle=none,fillstyle=solid,fillcolor=white](-2,-1)(2,1)
\pscircle[fillstyle=solid,fillcolor=red](-3,-1){0.1}
\psline[linewidth=2pt,linecolor=gray](-2.4,-1.4)(-1.6,-0.6)
\psline[linewidth=2pt,linecolor=gray](1.6,0.6)(2.4,1.4)
\psline[linewidth=2pt,linecolor=black](-2.4,0.6)(-1.6,1.4)
\psline[linewidth=2pt,linecolor=black](1.6,-1.4)(2.4,-0.6)
\psline[linecolor=red]{->>}(-3,-1)(-2,-1)
\psline[linecolor=red]{->}(-2,-1)(2,-1)
\psline[linecolor=red]{->}(2,-1)(2,1)
\psline[linecolor=red]{->}(-2,-1)(-2,1)
\psline[linecolor=red]{->}(-2,1)(2,1)
\psline[linecolor=red]{->>}(2,1)(3,1)}}
\pstPlanePut[plane=xy,planecorr=normal](0,0,7){\rput{70}(0,0){
\psframe[linestyle=none,fillstyle=solid,fillcolor=white](-2,-1)(2,1)
\pscircle[fillstyle=solid,fillcolor=red](-3,-1){0.1}
\psline[linewidth=2pt,linecolor=gray](-2.4,-1.4)(-1.6,-0.6)
\psline[linewidth=2pt,linecolor=gray](1.6,0.6)(2.4,1.4)
\psline[linewidth=2pt,linecolor=black](-2.4,0.6)(-1.6,1.4)
\psline[linewidth=2pt,linecolor=black](1.6,-1.4)(2.4,-0.6)
\psline[linecolor=red]{->>}(-3,-1)(-2,-1)
\psline[linecolor=red]{->}(-2,-1)(2,-1)
\psline[linecolor=red]{->}(2,-1)(2,1)
\psline[linecolor=red]{->}(-2,-1)(-2,1)
\psline[linecolor=red]{->}(-2,1)(2,1)
\psline[linecolor=red]{->>}(2,1)(3,1)}}
\pstPlanePut[plane=xy,planecorr=normal](0,0,8){\rput{80}(0,0){
\psframe[linestyle=none,fillstyle=solid,fillcolor=white](-2,-1)(2,1)
\rput(3,1.2){analyzer}
\pscircle[fillstyle=solid,fillcolor=red](-3,-1){0.1}
\psline[linewidth=2pt,linecolor=gray](-2.4,-1.4)(-1.6,-0.6)
\psline[linewidth=2pt,linecolor=gray](1.6,0.6)(2.4,1.4)
\psline[linewidth=2pt,linecolor=black](-2.4,0.6)(-1.6,1.4)
\psline[linewidth=2pt,linecolor=black](1.6,-1.4)(2.4,-0.6)
\psline[linecolor=red]{->>}(-3,-1)(-2,-1)
\psline[linecolor=red]{->}(-2,-1)(2,-1)
\psline[linecolor=red]{->}(2,-1)(2,1)
\psline[linecolor=red]{->}(-2,-1)(-2,1)
\psline[linecolor=red]{->}(-2,1)(2,1)
\psline[linecolor=red]{->>}(2,1)(3,1)}}
\end{pspicture}}
\end{center}
\caption{A rotating and accelerating Mach--Zehnder interferometer defines a worldtube. Each constituent of the interferometer, e.g., the beam splitter, the mirrors, the analyzer, moves on a 4--worldline in $\mathscr{M}$. In a stationary space-time each of these worldlines are one point in the quotient space $\mathscr{N}$ which in this Figure is isomorphic to the $x-y$--plane. \label{Fig:4DInterferometer}}
\end{figure}
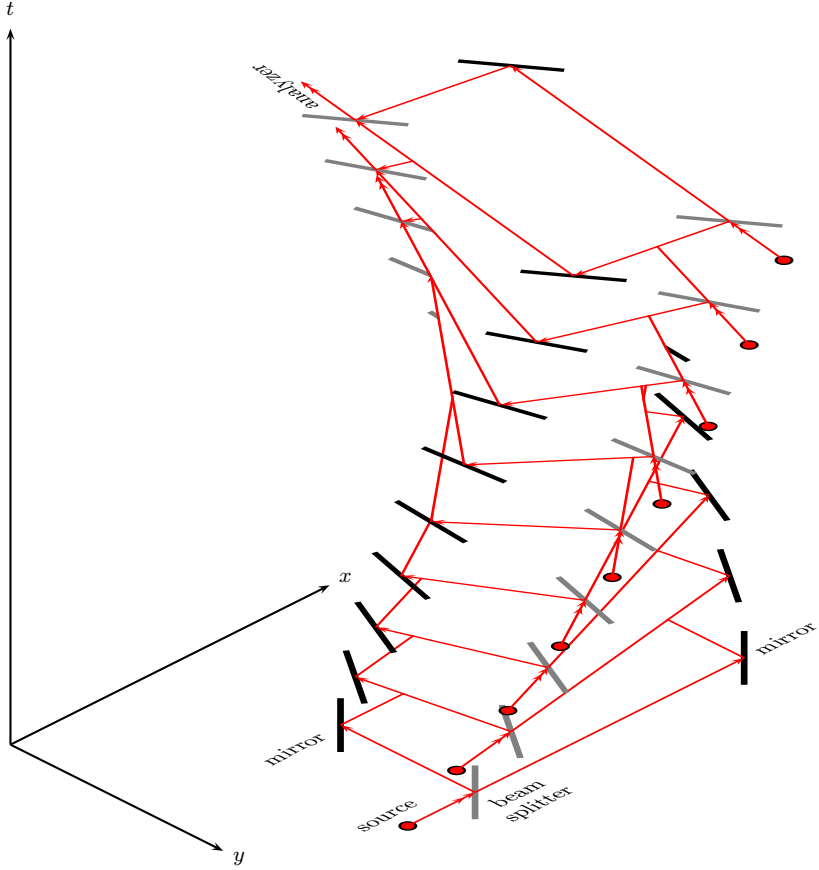

In this article we describe the interference of a scalar field in a Mach--Zehnder type interferometer. The outline of the paper is as follows: First we solve the Klein--Gordon equation in a general Riemannian space--time within a quasiclassical approximation. Then we present the mathematical description of the Mach--Zehnder interferometer which has been realized by neutron as well as atomic interferometry. In relativistic terms such a Mach--Zehnder interferometer forms a worldtube, see Fig.~\ref{Fig:4DInterferometer}. The combination of the propagating wave and the interferometer geometry results in the observable phase shift. This phase shift is then specialized to a Pleba\'{n}ski--Demia\'{n}ski space-time which depends on seven parameters. We discuss how the various parameters of the metric contribute to the calculated phase shift. This represents a generalization of the description of interferometery of neutral particles in this class of space--times \cite{KagramanovaLaemmerzahl07}. In our approach we put particular emphasis on a gauge invariant formalism for the description of space--time symmetries of the electromagnetic field. This formalism naturally leads to, e.g., gravitationally modified gauge invariant electrostatic and magnetostatic potentials. These $U(1)$ gauge invariant potentials in turn define a restricted gauge freedom which is uniquely related to the choice of a scale and a zero of the energy. This describes the fact that experimentally only ratios of differences of energies can be measured. 

\section{Dynamics of the matter field}

We take as quantum system a scalar field with electric and magnetic charge. This scalar field is assumed to obey the Klein--Gordon equation minimally coupled to the electromagnetic field and to gravitation
\begin{equation}
g^{\mu\nu} \mathscr{D}_\mu \mathscr{D}_\nu \psi - m^2 \psi = 0 \, .
\end{equation}
Here $g_{\mu\nu}$ is the non--singular space--time metric with signature $+2$ and $g^{\mu\nu}$ its inverse defined by $g^{\mu\rho} g_{\rho\nu} = \delta^\mu_\nu$. We also set $\hbar = c = 1$; greek indices run from $0$ to $3$. The covariant derivative $\mathscr{D}_\mu$ is defined by $\mathscr{D}_\mu T^\nu = D_\mu T^\nu - i e_p A_{\mu}T^\nu - g_p\check{A}_{\mu}T^\nu $, with $D_\mu T^\nu = \partial_\mu T^\nu + \left\{\begin{smallmatrix}\nu \\ \mu\rho\end{smallmatrix}\right\} T^\rho$ where $\left\{\begin{smallmatrix}\mu \\ \nu\rho\end{smallmatrix}\right\} = \frac{1}{2} g^{\mu\sigma} \left(\partial_\nu g_{\rho\sigma} + \partial_\rho g_{\nu\sigma} - \partial_\sigma g_{\nu\rho}\right)$ is the Christoffel symbol and $e_p$ and $g_p$ are the electric and magnetic charges of the field $T^{\mu}$. The Maxwell potential $A_\mu$ gives the Maxwell field strength $F_{\mu\nu} = \partial_\mu A_\nu - \partial_\nu A_\mu$. In vacuum $\partial_\mu F^{\mu\nu} = 0$ and $\partial_\mu \check F^{\mu\nu} = 0$ where $\check F^{\mu\nu} := \frac{i}{2 \sqrt{-g}} \epsilon^{\mu\nu\rho\sigma} F_{\rho\sigma}$ is the dual field strength and $\epsilon^{\mu\nu\rho\sigma}$ the Levi--Civita symbol. The dual potential $\check{A}_\mu$ is related to the dual field strength, see e.g. \cite{Plebanski75}. Note that $\check{A}$ and $\check{F}$ are purely imaginary. 

The solution of the Klein--Gordon equation we are looking for can be obtained through the ansatz $\psi(x) = a(x) e^{i S(x)}$ where $S$ is assumed to be real valued and $a$ to be invariant under gauge transformations. We insert this ansatz into the field equation and obtain
\begin{align}
0 & =  g^{\mu\nu} \mathscr{D}_\mu \mathscr{D}_\nu \left(a e^{i S}\right) - m^2 a e^{i S} \nonumber\\
& =  g^{\mu\nu} \left[(D_\mu D_\nu a e^{i S} + 2 i D_\nu a \left(\partial_\mu S - e_p A_\mu + i g_p \check A_\mu\right) e^{i S} + a i D_\mu \left(\partial_\nu S - e_p A_\nu + i g_p \check A_\nu\right) e^{i S} \right. 
\nonumber\\
&  \left. - a \left(\partial_\nu S - e_p A_\nu + i g_p \check A_\nu\right) \left(\partial_\mu S - e_p A_\mu + i g_p \check A_\mu\right) e^{i S}\right] - m^2 a e^{i S}
\end{align}
With $\mathscr{P}_\mu = \partial_\mu S$ and $p_\mu = \mathscr{P}_\mu - e_p A_\mu + i g_p \check{A}_\mu$ this reads
\begin{eqnarray}
0 & = & g^{\mu\nu} D_\mu D_\nu a + i g^{\mu\nu} \left(2 D_\nu a p_\mu + D_\mu p_\nu a\right) - g^{\mu\nu} p_\nu p_\mu a - m^2 a \, . \label{KGAnsatz}
\end{eqnarray}

Now we assume that the amplitude $a$ and the kinetic momentum $p_\mu$ vary very slowly compared to the variation of the phase $S$. That means
\begin{equation}
\partial_\mu a \ll p_\mu a \, , \qquad g^{\mu\nu} D_\mu p_\nu \ll m^2 \, . \label{WKBConditions}
\end{equation}
These are the conditions for the quasiclassical approximation. If we take $\hbar \neq 1$ then these conditions can be mimicked by the requirement that $\hbar$ is small. We will come back to these approximation conditions.

With \eqref{WKBConditions} we obtain from \eqref{KGAnsatz} to lowest order
\begin{equation}
0 = \left(- g^{\mu\nu} p_\nu p_\mu - m^2\right) a \, ,
\end{equation}
that is,
\begin{equation}
g^{\mu\nu} p_\mu p_\nu = - m^2 \, . \label{HamJac}
\end{equation}
This is the general relativistic Hamilton--Jacobi equation for a point particle with mass $m$. Therefore $p_\mu$ and $\mathscr{P}_\mu$ can be interpreted as kinetic and canonical momentum, respectively. From the one--form $p_\mu$ we obtain the 4--velocity vector through $v^\mu = \frac{1}{m} g^{\mu\nu} p_\nu$. The integral curves of $v$ are realized by the trajectories of the peaks of wave packets. Differentiation of \eqref{HamJac} gives the Lorentz--force equation for a particle with electric and magnetic charge
\begin{equation}
m D_v v = e_p F(v) - i g_p \check{F}(v) \ . \label{geodesics}
\end{equation}

If \eqref{HamJac} is fulfilled then we obtain from \eqref{KGAnsatz}
\begin{equation}
0 = g^{\mu\nu} D_\mu D_\nu a - i \left(2 g^{\mu\nu} p_\mu D_\nu a + g^{\mu\nu} D_\mu p_\nu a\right)\, .
\end{equation}
Owing to the higher derivative the first term is smaller than the rest so that we obtain as first order approximation
\begin{equation}
0 = 2 g^{\mu\nu} p_\mu D_\nu a + g^{\mu\nu} D_\mu p_\nu a \, .
\end{equation}
We replace the momentum by the velocity and obtain
\begin{equation}
D_v a = - \frac{1}{2} \theta a \, ,
\end{equation}
where $\theta = D_\mu v^\mu$ which is interpreted as expansion of the congruence of worldlines given by the integral curves of $v$. This is a propagation equation for the amplitude $a$. This equation can be rewritten as
\begin{equation}
\frac{1}{a} \frac{d}{ds} a = \frac{d}{ds} \ln a = - \frac{1}{2} \theta
\end{equation}
with the solution
\begin{equation}
a = \exp\left(- \frac{1}{2} \int \theta ds\right) a_0
\end{equation}
where $a_0 = a(x_0)$ with the initial point $x_0$.

The quasiclassical wave function then is given by
\begin{equation}
\psi(x) = \exp\left(- \frac{1}{2} \int_{x_0}^x \theta ds\right) \exp\left(- i \int_{x_0}^x \mathscr{P}_\mu dx^\mu\right) a_0
\end{equation}
where the integration is along the trajectories given by the solutions of \eqref{geodesics}.

Now we come back to the approximation condition \eqref{WKBConditions}. The second condition implies $\theta \ll m = 1/\lambda_{\rm C}$, where $\lambda_{\rm C}$ is the Compton wavelength. In other words, $\lambda_{\rm C} \theta \ll 1$. The expansion $\theta$ characterizes the change of the spatial distance between two neighboring trajectories: if $\delta r$ denotes a spacelike vector between two neighboring trajectories then $D_u \ln\rho = \frac{1}{3}\theta$ where $\rho^2 = g(\delta r, \delta r)$. As a consequence our condition reads $\lambda_{\rm C} D_u \ln\rho \ll 1$ and means that the relative change of the distance between neighboring trajectories is very small within a time span of a Compton time. Using the Raychaudhury relation this also implies a condition on the strength of the space--time curvature and the electromagnetic field.  

\section{Model of the interferometer}

\subsection{Stationary congruences}

A congruence of Killing trajectories is a set of worldlines whose tangent vector is proportional to a Killing vector field. A Killing field $\xi$ fulfills $0 = \mathscr{L}_\xi g$ where $\mathscr{L}_\xi$ is the Lie derivative with respect to $\xi$. This is equivalent to $D_\mu \xi_\nu + D_\nu \xi_\mu = 0$.

If $u$ is a normalized vector field proportional to $\xi$ we have
\begin{equation}
\xi = e^U u \, , 
\end{equation}
where $g(u,u) = -1$. The exponent $U$ is a generalized gravitational potential since its gradient gives the acceleration of the Killing trajectory, $a = D_u u = g(\cdot, d U)$. It is easy to show that this potential is constant along the Killing trajectories, $D_u U = 0$. The generalized potential $U$ combines gravitational as well as inertial forces.

We also need the projector $P_\xi$ defined through 
\begin{equation}
P_\xi A = A - \frac{g(A, \xi)}{g(\xi, \xi)} \xi  \, ,
\end{equation}
which projects a vector $A$ onto the rest space of $u$. It is $P_\xi P_\xi = P_\xi$ and $P_\xi = P_u$.

The rotation $\omega_{\mu\nu}$ of a congruence is defined by $\omega_{\mu\nu} = (P_u)^\rho_\mu (P_u)^\sigma_\nu D_{[\rho} u_{\sigma]}$. In terms of the Killing field we obtain \cite{Ehlers61}
\begin{equation}
\omega_{\mu\nu} = - e^{-U} D_{[\mu} \xi_{\nu]} + 2 \partial_{[\mu}U u_{\nu]} \, .
\end{equation}
Note that $\omega_{\mu\nu} \xi^\nu = 0$ and $\mathscr{L}_\xi \omega = 0$. The stationarity of a congruence is equivalent to three conditions which have a direct physical interpretation: (i) the congruence is rigid, (ii) the rotation vector in the rest frame does not precess, and (iii) the acceleration in the rest frame of the congruence is co--rotating \cite{Ehlers61}.  

The existence of Killing trajectories in a 4--dimensional manifold $\mathscr{M}$ allows to establish a 3--dimensional quotient space $\mathscr{N}$ with an induced positive definite metric $h_{\mu\nu}$ \cite{Ehlers62,Geroch71}. This metric on $\mathscr{N}$ and its inverse are given by
\begin{equation}
h_{\mu\nu} = g_{\mu\nu} - \frac{1}{g(\xi,\xi)} \xi_\mu \xi_\nu \, , \qquad h^{\mu\nu} = g^{\mu\nu} - \frac{1}{g(\xi,\xi)} \xi^\mu \xi^\nu \, .
\end{equation}
In adapted coordinates the covariant form of the metric tensor adapts the form $h_{ij} = g_{ij} - g_{0i} g_{0j}/g_{00}$ \cite{AbramowiczCarterLasota88}.

\subsection{Stationarity of the electromagnetic field}

In addition to the stationarity of the interferometer we also require the stationarity of the electromagnetic field given by $\mathscr{L}_\xi F=0$. The Lie derivative acting on differential forms employs the identity $\mathscr{L}_\xi = i_\xi d + d i_\xi$, where $d$ is the exterior derivative operator and $i_\xi$ is the interior product. With the homogeneous Maxwell equations $\mathscr{L}_\xi F=0$ this implies that $\mathscr{L}_\xi A = d\lambda$ for some function $\lambda$ (here one uses $\mathscr{L}_\xi d = d \mathscr{L}_\xi$). Since for Killing vectors $\mathscr{L}_\xi$ commutes with the duality operation we also have $\mathscr{L}_\xi\check{F} = 0$. With the Maxwell equations in vacuum this yields $\mathscr{L}_\xi\check{A} = d\check{\lambda}$ for some function $\check{\lambda}$. This means that the potentials are stationary up to a gauge transformation.

From $0 = \mathscr{L}_\xi F = i_\xi d F + d i_\xi F = d i_\xi F$ we also infer that there is a function $\phi$ so that $i_\xi F = -d\phi$. The function $\phi$ is a generalized electrostatic potential. Similarly, there is a generalized magnetostatic potential $\check{\phi}$ so that $i_\xi \check{F} = - d\check{\phi}$,
\begin{equation}
i_\xi F = - d\phi \, , \qquad i_\xi \check{F} = - d\check{\phi} \, . \label{GenElectricMagneticPotentials}
\end{equation}
It is obvious that these two generalized potentials are constant along the Killing trajectories, $\mathscr{L}_\xi \phi = 0$ and $\mathscr{L}_\xi \check{\phi} = 0$. It is also clear that $\phi$ and $\check{\phi}$ are unique up to a constant. 

From the generalized Lorentz force equation $D_v p = - e_p i_v  F + i g_p i_v \check{F}$ we obtain $(D_v p)(\xi) = D_v(p(\xi)) - p(D_v \xi) = e_p i_v i_\xi F - i g_p i_v i_\xi \check{F} = - e_p v(\phi) + i g_p v(\check{\phi})$. Since $\xi$ is a Killing vector $p(D_v \xi) = 0$. This gives a conserved quantity
\begin{equation}
\mathscr{E} = p(\xi) + e_p \phi - i g_p \check{\phi} = const \, , \label{ConservedEnergy}
\end{equation}
which is interpreted as total conserved energy consisting of a gravitationally modified kinetic and rest energy $p(\xi)$, a  modified electrostatic energy $e_p \phi$ and a modified magnetostatic energy $ig_p \check{\phi}$. One may define a non--gravitational potential energy $\mathscr{E}_{\rm pot} = e_p \phi - i g_p \check{\phi}$. Then $\mathscr{E} = p(\xi) + \mathscr{E}_{\rm pot}$. Note that the modified electrostatic and magnetostatic potentials are manifestly gauge invariant. 

The function $\phi$ (the same holds for the function $\check{\phi}$) is a generalized electrostatic (magnetostatic) potential. Choosing an observer $u$ collinear with the Killing field then $d\phi = e^U E$ where $E = i_u F$ is the electric field seen by the observer. 
From $- d\phi = i_\xi F = i_\xi dA = \mathscr{L}_\xi A - d i_\xi A$ it is clear that in a gauge given by $\mathscr{L}_\xi A = 0$ we have $\phi = A(\xi)$.

The Killing vector is defined up to a constant factor. The generalized electrostatic and magnetostatic potentials are defined only up to an additive constant. These two degrees of freedom allow the transformations 
\begin{equation}
\xi \rightarrow \xi^\prime = \alpha \xi \, , \qquad \phi \rightarrow \phi^\prime = \phi + \beta \, , \qquad \check{\phi} \rightarrow \check{\phi}^\prime = \check{\phi} + i \check{\beta} \, ,\label{NewGaugeTransformations}
\end{equation}
where $\alpha, \beta, \check{\beta} \in \mathbbm{R}$ with $\alpha \neq 0$. As a consequence, the total energy $\mathscr{E}$ transforms as  
\begin{equation}
\mathscr{E} \rightarrow \mathscr{E}^\prime = \alpha \mathscr{E} + \alpha \left(e_p \beta + g_p \check{\beta}\right) \, .  \label{EnergyTransformations}
\end{equation}
Therefore, the factor $\alpha$ rescales the energy and $\alpha \left(e_p \beta + g_p \check{\beta}\right)$ selects a new zero. This is in agreement with the
experimental procedure of energy measurement where always differences of energies are compared -- and such ratios of energy differences are in fact invariant under the affine transformations \eqref{EnergyTransformations}.

\subsection{The model of the interferometer}

We assume that our interferometer is stationary and that the matter wave leaving the source is stationary, too. In more mathematical terms:
\begin{enumerate}
\item The stationarity of the interferometer means that the worldline of each constituent of the interferometer is a Killing trajectory. Therefore, the 4--velocity of each constituent $u$ is proportional to a Killing vector $\xi$, $u \sim \xi$.
\item The stationarity of the source means that the kinetic momentum of the matter wave emitted by the source is time--independent with respect to the interferometer frame, $\left.\partial_0 p\right|_{\rm source} \stackrel{*}{=} 0$, where $\stackrel{*}{=}$ denotes equality at the source in that particular frame.
\end{enumerate}
We now draw some consequences from these assumptions.

\subsection{Consequences}

The condition of the stationarity of the source is equivalent to $\left.\mathscr{L}_\xi p\right|_{\rm source} = 0$. Since in a stationary space--time the covariant derivative and the Lie derivative commute, we have $\mathscr{L}_\xi p = 0$ everywhere. The latter also implies $\mathscr{L}_\xi \theta = 0$. 
For the Lie--derivative of the canonical momentum we obtain $\mathscr{L}_\xi \mathscr{P} = - e_p d\lambda + i g_p d\check{\lambda}$. This implies that the closed loop integral of the canonical momentum, the phase shift to be derived below, is time independent:
\begin{equation}
\frac{d}{dt} \oint \mathscr{P} = \oint \mathscr{L}_\xi \mathscr{P} = - e_p \oint d\lambda + i g_p \oint d\check{\lambda} = 0 \ .
\end{equation}

The momentum in the rest space of the interferometer is given by $P_u p$. This 1--form on $\mathscr{M}$ satisfies $(P_u p)(\xi) = 0$ and $\mathscr{L}_\xi P_u p = 0$. Therefore, $P_u p$ is also a 1--form in $\mathscr{N}$ \cite{Geroch71} which we denote by $\mbox{\boldmath$p$}$. The modulus of the momentum measured in $\mathscr{N}$ is $|\mbox{\boldmath$p$}|^2 = h(\mbox{\boldmath$p$}, \mbox{\boldmath$p$})$ \cite{Geroch71}. It follows from the dispersion relation \eqref{HamJac}
\begin{equation}
|\mbox{\boldmath$p$}| = \sqrt{(p(\xi))^2 e^{- 2 U} - m^2} = \sqrt{\left(\mathscr{E} - \mathscr{E}_{\rm pot}\right)^2 e^{- 2 U} - m^2} \ . \label{modulusmomentum}
\end{equation}
Since $\mathscr{L}_\xi U = 0$, $\mathscr{L}_\xi \phi = 0$, and $\mathscr{L}_\xi \check{\phi} = 0$ the functions $U$, $\phi$, and $\check{\phi}$ are also scalar functions on $\mathscr{N}$.

Note that $\omega_{\mu\nu}$ is a 2--form on the quotient space $\mathscr{N}$ and, thus, will be denoted by $\mbox{\boldmath$\omega$}$.

\section{Pleba\'{n}ski--Demia\'{n}ski space--time}

The initial form of the Pleba\'{n}ski--Demia\'{n}ski metric in real coordinates describing the space--time with nonzero cosmological constant of a rotating and accelerating source endowed with mass, electric, magnetic and gravitomagnetic charges can be found in \cite{PlebanskiDemianski76}. We follow the coordinate transformations presented in \cite{HongTeo05,GriffithsPodolsky06}
and write the metric of generalized accelerating black holes in the form
\begin{eqnarray}
\label{metric_non_accelerating}
ds^2 & = & \frac{1}{\Omega^2}\left(- \frac{\Delta}{\rho^2} \left(dt - (a \sin^2\theta + 2 q (1 - \cos\theta)) d\varphi\right)^2 + \frac{\rho^2}{\Delta} dr^2
\right. \nonumber\\
& & \left. + \frac{\bar{P}}{\rho^2}\left(a dt - (r^2 + (a+q)^2) d\varphi\right)^2
+\rho^2\frac{\sin^2\theta}{\bar{P}}d\theta^2 \right ) \ ,
\end{eqnarray}
with
\begin{eqnarray}
\rho^2 & = & r^2 + (q + a \cos\theta)^2 \\
\Omega & = & 1 - \frac{\alpha}{w} (q + a \cos\theta) r \\
\bar{P} & = & \sin^2\theta \left(1 - a_3\cos\theta - a_4\cos^2\theta \right) \\
\Delta & = & (\kappa + e^2 + g^2) - 2 m r + \epsilon r^2 - 2 n \frac{\alpha}{w}r^3 - \left(\frac{\alpha^2}{w^2} \kappa + \frac{\Lambda}{3}\right) r^4 
\end{eqnarray}
where
\begin{eqnarray}
a_3 & = & 2 a \frac{\alpha}{w} m - 4 a q \frac{\alpha^2}{w^2} (\kappa + e^2 + g^2) - 4\frac{\Lambda}{3}aq \\
a_4 & = & - a^2 \frac{\alpha^2}{w^2}(\kappa + e^2 + g^2)-\frac{\Lambda}{3}a^2\\
\epsilon & = & \frac{\kappa}{a^2-q^2} + 4 q \frac{\alpha}{w} m - (a^2 + 3 q^2) \left( \frac{\alpha^2}{w^2} (\kappa + e^2 + g^2) + \frac{\Lambda}{3} \right) \\
n & = & \frac{\kappa q}{a^2-q^2} - (a^2 - q^2) \frac{\alpha}{w} m + (a^2 - q^2) q \left(\frac{\alpha^2}{w^2} (\kappa + e^2 + g^2) + \frac{\Lambda}{3} \right) \\
\kappa & = & \frac{1 + 2 q \frac{\alpha}{w}m - 3 q^2 \frac{\alpha^2}{w^2}(e^2+g^2) - q^2 \Lambda}{\frac{1}{a^2-q^2} + 3 q^2 \frac{\alpha^2}{w^2}}  \ .
\end{eqnarray}
We have
\begin{align}
g_{00} & =  \frac{- \Delta + \bar P  a^2}{\Omega^2 \rho^2} \\
g_{0i} dx^i & =  \frac{1}{\Omega^2}\left[ \frac{\Delta}{\rho^2} \left(a \sin^2\theta + 2 q (1 - \cos\theta)\right) - \frac{\bar{P}}{\rho^2} a \left(r^2 + (a + q)^2\right)\right] d\varphi \\
g_{ij} dx^i dx^j & =  \frac{\rho^2}{\Omega^2\Delta} dr^2 + \rho^2 \frac{\sin^2\theta}{\Omega^2\bar{P}} d\theta^2 + \frac{1}{\Omega^2} \left[- \frac{\Delta}{\rho^2} \left(a \sin^2\theta + 2 q (1 -
\cos\theta)\right)^2 \right. \nonumber \\
&  \left. \qquad\qquad\qquad\qquad\qquad\qquad + \frac{\bar{P}}{\rho^2} \left(r^2 + (a + q)^2\right)^2\right] d\varphi^2 \ .
\end{align}
This family of solutions are characterized by a mass--like parameter $m$, a cosmological constant $\Lambda$, a rotation--like parameter $a$, a NUT--like parameter $q$, the electric and magnetic charges $e$ and $g$, an acceleration--like parameter $\alpha$, and the twist parameter $w$. The latter can be given any convenient value provided both $a$ and $q$ are nonvanishing (all limits of this family of solutions including the limit of vanishing $a$ or $q$ and the corresponding choice of the parameter $w$ can be found in \cite{GriffithsPodolsky06}).

In order to obtain measurable components of physical quantities one has to introduce a tetrad defined by $e_{\hat{t}}=u$ and $e_{\hat{a}} e_{\hat{b}} = \eta_{\hat{a}\hat{b}}$, where $\eta_{\hat{a}\hat{b}}= {\rm diag}(-1,1,1,1)$. Henceforth tetrad components of a quantity are denoted by symbol  \  $\hat{}$\phantom{l}.   
The components of the tetrad frame for the stationary observer are
\begin{align} 
e_{\hat t}^\mu & = \frac{1}{\sqrt{-g_{00}}}\bigg(1,0,0,0\bigg) \, , \qquad & e^{\hat t}_\mu & = -\sqrt{-g_{00}}\bigg(1,0,0,\frac{g_{0\varphi}}{g_{00}} \bigg)  \label{tetrad_0} \\
e_{\hat r}^\mu & = \frac{1}{\sqrt{g_{rr}}}\bigg(0,1,0,0\bigg) \,  ,  \qquad  & e^{\hat r}_{\mu} & = \sqrt{g_{rr}}\bigg(0,1,0,0\bigg) \label{tetrad_1} \\
e_{\hat\theta}^\mu  & = \frac{1}{\sqrt{g_{\theta\theta}}}\bigg(0,0,1,0\bigg)  \, ,  \qquad  & e^{\hat\theta}_\mu & = \sqrt{g_{\theta\theta}}\bigg(0,0,1,0\bigg) \label{tetrad_2}  \\
e_{\hat\varphi}^\mu & = \sqrt{\frac{-g_{00}}{g_{0\varphi}^2 - g_{00}g_{\varphi\varphi}}} \bigg(-\frac{g_{0\varphi}}{g_{00}},0,0,1\bigg) \, , \qquad & e^{\hat \varphi}_\mu & = \sqrt{\frac{g_{0\varphi}^2 - g_{00}g_{\varphi\varphi}}{- g_{00}}}\bigg(0,0,0,1\bigg) \ . \label{tetrad_3}
\end{align} 

Since all metrical coefficients do not depend on $t$, this metric admits a Killing vector $\xi^\mu = \delta^\mu_0$. Therefore,
\begin{equation}
-e^{2 U} = g(\xi, \xi) = g_{00} = \frac{- \Delta + \bar P  a^2}{\Omega^2 \rho^2} 
\end{equation}
and the acceleration of the Killing trajectories is 
\begin{equation}
a_\mu = \partial_\mu U = \frac{1}{2} \partial_\mu \ln (-g_{00}) =  \frac{1}{2} \partial_\mu \ln\left( \frac{\Delta - \bar P  a^2}{\Omega^2 \rho^2}\right) \, .
\end{equation}
Though the acceleration and the rotation of the Killing vector field can be calculated exactly, for the application to a realistic experimental situation on Earth the restriction to weak fields is sufficient and also leads to much simpler results. Therefore we assume the parameters $m$, $\Lambda$, $a$, $q$, $\alpha$, $w$, $e$
and $g$ to be small and expand the metric coefficients to that order where these parameters appear first (e.g., first order in $\Lambda$ and second order in $e$ since there is no first order term for $e$). We obtain
\begin{eqnarray}
\hspace{-0.5cm} g_{00} & \approx & - 1 + \frac{2 m}{r} + \frac{\Lambda}{3} r^2 + \frac{1}{r^2} \left(2 q(q + a \cos\theta) - e^2 - g^2 \right) -2 r a \frac{\alpha}{w} \cos\theta \ .
\end{eqnarray}
and
\begin{eqnarray}
a_{\hat{r}} & \approx & \frac{m}{r^2} - \frac{\Lambda}{3}(r+m) + \frac{1}{r^3} \left(2 q(q + a \cos\theta) - e^2 - g^2 \right) + a \frac{\alpha}{w} \cos\theta  \label{acceleration1} \\
a_{\hat{\theta}} & \approx & \frac{a q \sin\theta}{r^3} - a \frac{\alpha}{w} \sin\theta \label{Killing0acc2}  \\
a_{\hat{\varphi}} & = & 0 \ .
\end{eqnarray}
(It can be seen from \eqref{acceleration1} that there are two constant terms in $a_{\hat r}$. Both terms have recently been discussed as to whether they can be a possible origin of the Pioneer anomaly \cite{BiniCherubiniMashhoon04,KagramanovaKunzLaemmerzahl06}.)

The rotation of the stationary congruence $\omega_{\mu\nu}$ in the Pleba\'{n}ski--Demia\'{n}ski space--time is given by
\begin{eqnarray}
\omega_{\hat{r}\hat{\theta}} & = & 0  \\
\omega_{\hat{r}\hat{\varphi}} & = & a \left(\frac{\Lambda}{3} - \frac{m}{r^3} + \frac{q}{r} \frac{\alpha}{w}\right) \sin\theta  \label{Killing0rot2} \\
\omega_{\hat{\theta}\hat{\varphi}} & = & - \frac{q}{r^2} + \left(\frac{\Lambda}{6} + \frac{m}{r^3}\right)(q + 2 a \cos\theta) + \frac{q a}{r} \frac{\alpha}{w} \cos\theta \ . \label{rotation3}
\end{eqnarray}
If we regard the acceleration as the ``electric'' and the rotation as the ``magnetic'' part of the gravitational field, then the above expressions give rise to the interpretation that $m$ is the gravitoelectric mass and $q$ the gravitomagnetic mass. The product $a m$ may be regarded as a gravitomagnetic moment. 

Introducing the electric $A_{\mu}$ and magnetic $\check{A}_{\mu}$ vector potentials  for the metric \eqref{metric_non_accelerating} with components
\begin{align}
A_{\varphi}&=-\frac{re}{\rho^2}\left( a\sin^2\theta + 2 q (1-\cos\theta) \right) - \frac{g}{a\rho^2}(q+a\cos\theta)(r^2+(q+a)^2)
\ , \qquad \nonumber \\
A_{t}&=\frac{er + (q+a\cos\theta)g}{\rho^2} \ , \qquad \check{A}_{t}=i\frac{gr - (q+a\cos\theta)e}{\rho^2} \ , \nonumber \\
\check{A}_{\varphi}&=-\frac{ig}{\rho^2}\left( a\sin^2\theta + 2 q (1-\cos\theta) \right) + \frac{ie}{a\rho^2}(q+a\cos\theta)(r^2+(q+a)^2) \ ,
\end{align}
we can write the total energy of the particle in the weak field approximation in the following form 
\begin{equation}
\mathscr{E} = p(\xi) + \mathscr{E}_{pot} = p(\xi) + e_p A_t - i g_p \check{A}_t = p(\xi) + \frac{e_p e + g_p g}{r} + \frac{(e_p g - g_p e)(q+a\cos\theta)}{r^2} \, ,
\end{equation}
where $e_p$ and $g_p$ are electric and magnetic charges of the particle. 

For later use we note the measured components of the electromagnetic field
\begin{align}
B_{\hat{r}} & = \frac{g}{r^2} - 2\left(q+a\cos\theta\right) \left(\frac{\alpha}{w}\frac{g}{r} + \frac{e}{r^3}\right)  \, , & \qquad 
B_{\hat{\theta}} & = -e\frac{a}{r^2} \left(\frac{1}{r} - q \frac{\alpha}{w}  \right) \sin\theta  \label{B_Field} \ , \\
E_{\hat{r}} & = -\frac{e}{r^2} + 2\left(q+a\cos\theta\right) \left(\frac{\alpha}{w}\frac{e}{r}-\frac{g}{r^3} \right)  \, ,  & \qquad 
E_{\hat{\theta}} & = -g \frac{a}{r^2} \left(\frac{1}{r} - q \frac{\alpha}{w}  \right) \sin\theta \,  . \label{E_Field}
\end{align}

\section{The phase shift}

\subsection{The general expression}

The wave function propagates along two orbits, I and II, both starting at $x_0$ and ending at $x$, see also Fig.~\ref{Fig:Interferometer}. The wave functions at those points are
\begin{eqnarray}
\psi_{\rm I}(x) & = & \exp\left(- \frac{1}{2} \int_{x_0}^x \!\!\!\!\!\!\!\!\!\!\circlearrowleft \theta ds\right) \exp\left(- i \int_{x_0}^x\!\!\!\!\!\!\!\!\!\!\circlearrowleft \mathscr{P}\right) a_0  \label{phiI}\\
\psi_{\rm II}(x) & = & \exp\left(- \frac{1}{2} \int_{x_0}^x \!\!\!\!\!\!\!\!\!\!\circlearrowright \theta ds\right) \exp\left(- i \int_{x_0}^x\!\!\!\!\!\!\!\!\!\!\circlearrowright \mathscr{P}\right) a_0 \, ,\label{phiII}
\end{eqnarray}
where the upper integral is along path I and the lower along path II. The attenuation factor $\int^x \theta ds$ will not play any role. The reason is that this factor only describes the attenuation of the density of the worldlines given by the phase $S$. However, no particle will disappear since we still have current conservation. At the analyzer the total intensity will be measured which is an integral over the probability density. This adds up to unity since nothing can disappear. Therefore we can omit the consideration of the attenuating factor.

The intensity of the two interfering wave functions at one port of the analyzer is
\begin{equation}
I = |\psi_{\rm I} + \psi_{\rm II}|^2 = 2 \left(1 + \cos\Delta\Phi\right) |a_0|^2 \, ,
\end{equation}
where according to \eqref{phiI} and \eqref{phiII} the phase shift $\Delta\Phi$ is given by
\begin{equation}
\Delta\Phi = \oint \mathscr{P} = \oint p + \oint \left(e_p A - i g_p \check{A}\right)  \, .
\end{equation}
The second term is an Aharonov--Bohm like phase shift and the third term its magnetic analoque. The first term gives the phase shift which stems from the change of the kinetic momentum due to the particle's interaction with the gravitational and electromagnetic field.

We decompose the kinetic term $p = p(u) u + P_u p = p(\xi) e^{- U} u + P_u p$ and introduce the conserved energy \eqref{ConservedEnergy}
\begin{equation}
\Delta\Phi = \mathscr{E} \oint e^{- U} u - \oint \left(e_p \phi - i g_p \check{\phi}\right) e^{- U} u + \oint P_u p + \oint \left(e_p A - i g_p \check{A}\right)  \, .
\end{equation}
Stoke's law gives
\begin{eqnarray}
\Delta\Phi & = & \mathscr{E} \int e^{- U} \omega - \int \left(\left(e_p d \phi - i g_p d \check{\phi}\right) \wedge e^{- U} u + \left(e_p \phi - i g_p \check{\phi}\right) e^{- U} \omega\right) 
\nonumber \\
& & \qquad\qquad + \oint P_u p + \int \left(e_p F - i g_p \check{F}\right) \, .
\end{eqnarray}
With \eqref{GenElectricMagneticPotentials} this yields
\begin{equation}
\Delta\Phi = \mathscr{E} \int e^{- U} \omega - \int \left(e_p \phi - i g_p \check{\phi}\right) e^{- U} \omega + \oint P_u p + \int P_u \left(e_p F - i g_p \check{F}\right) \, ,
\end{equation}
where we used the decomposition $F = P_u F + i_u F \wedge u$ (here $P_u$ acts on both indices of $F$). 
Since $P_u p$ is a 1--form and $\omega$ as well as $P_u F$ and $P_u \check{F}$ are 2--forms on the quotient space $\mathscr{N}$ we can write
\begin{equation}
\Delta\Phi = \mathscr{E} \int e^{- U} \mbox{\boldmath$\omega$} + \int \mathscr{E}_{\rm pot} e^{- U} \mbox{\boldmath$\omega$} + \oint \mbox{\boldmath$p$} + \int \left(e_p \mbox{\boldmath$F$} - i g_p \check{\mbox{\boldmath$F$}}\right) \, .
\end{equation}
where everything now is expressed in terms of the coordinates of $\mathscr{N}$. The last term describes the Aharonov--Bohm und dual Aharonov--Bohm effects~\cite{DowlingWilliamsFranson99,Furtado}.  
We introduce the magnetic and electric fluxes $\Phi_{\rm m} = \int \mbox{\boldmath$F$}$ and $\Phi_{\rm e} = i \int \check{\mbox{\boldmath$F$}}$ through the interferometer area. We also use \eqref{modulusmomentum} and the unit 1--form $\mbox{\boldmath$n$} = \mbox{\boldmath$p$}/|\mbox{\boldmath$p$}|$ in the direction of the momentum. Then the phase shift takes its final form
\begin{equation}
\Delta\Phi = \mathscr{E} \int e^{- U} \mbox{\boldmath$\omega$} + \int \mathscr{E}_{\rm pot} e^{- U} \mbox{\boldmath$\omega$} + \oint \sqrt{\left(\mathscr{E} - \mathscr{E}_{\rm pot}\right)^2 e^{-2 U} - m^2}\; \mbox{\boldmath$n$} + e_p \Phi_m - g_p \Phi_e \ , \label{phasegeneralFlux}
\end{equation} 
It should be noted that this phase shift is manifestly gauge invariant and also invariant under the transformations \eqref{NewGaugeTransformations}. 

The first two terms give a generalized Sagnac effect, the third term yields the phase shift due to the change of the potential energies over the interferometer, and the last two terms are Aharonov--Bohm like contributions. Here the Sagnac effect is modified by the gravitational potential and the non--gravitational potential energy. 

\subsection{Small interferometer}

\begin{figure}[t]
\begin{center}
{\footnotesize\begin{pspicture}(-4,-2)(4,2)
\rput(-3,-0.7){source}
\rput(-1.7,-1.5){\txt{beam \\ splitter}}
\rput(2.8,-1){mirror 2}
\rput(-2.9,1){mirror 1}
\rput(3,1.2){analyzer}
\pscircle[fillstyle=solid,fillcolor=red](-3,-1){0.1}
\psline[linewidth=2pt,linecolor=gray](-2.4,-1.4)(-1.6,-0.6)
\psline[linewidth=2pt,linecolor=gray](1.6,0.6)(2.4,1.4)
\psline[linewidth=2pt,linecolor=black](-2.4,0.6)(-1.6,1.4)
\psline[linewidth=2pt,linecolor=black](1.6,-1.4)(2.4,-0.6)
\psline[linecolor=red]{->>}(-3,-1)(-2,-1)
\psline[linecolor=red]{->}(-2,-1)(2,-1)
\psline[linecolor=red]{->}(2,-1)(2,1)
\psline[linecolor=red]{->}(-2,-1)(-2,1)
\psline[linecolor=red]{->}(-2,1)(2,1)
\psline[linecolor=red]{->>}(2,1)(3,1)
\psarc{<-}(-1,0){0.8}{90}{180}
\psarc{->}(1,0){0.8}{270}{360}
\rput(-1.2,0.2){I}
\rput(1.2,-0.2){II}
\psline{->}(-5,1)(-5,-1)
\rput(-5.5,0){$\nabla U$}
\psline{<-}(4.5,-1)(4.5,-0.2)
\psline{->}(4.5,0.2)(4.5,1)
\rput(4.5,0){$h$}
\psline{<-}(-2,-2)(-0.2,-2)
\psline{->}(0.2,-2)(2,-2)
\rput(0,-2){$l$}
\end{pspicture}}
\end{center}
\caption{The gradient of $U$ in the interferometer's rest frame. $h$ and $l$ are the interferometer's height and length. \label{Fig:Interferometer}}
\end{figure}
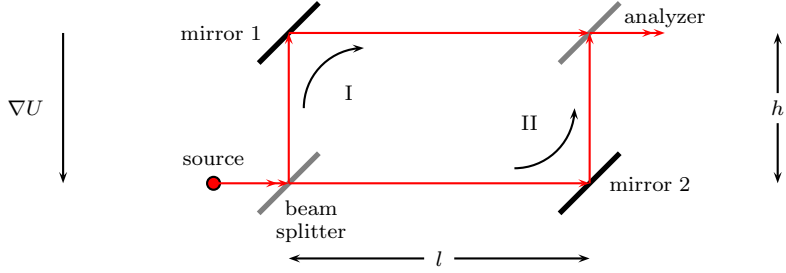

We now consider a small interferometer. Small interferometer means that all external fields like the gravitational potential as well as the electrostatic and magnetostatic potentials vary very slowly over the interferometer (Aharonov--Bohm like situations need a separate treatment). That means that the paths of the particles between the beam splitters, mirrors and analyzers are straight lines and the momenta of the particles are nearly constant. We also adjust the interferometer in such a way that the direction $\mbox{\boldmath$\nabla$} U$ is orthogonal to a line from the beam splitter to a mirror, see Fig.~\ref{Fig:Interferometer}. Then path I is from the beam splitter $\rightarrow$ mirror 1 $\rightarrow$ analyzer and path II from the beam splitter $\rightarrow$ mirror 2 $\rightarrow$ analyzer. As far as the third integral in \eqref{phasegeneralFlux} is concerned we obtain
\begin{align}
&\oint \sqrt{e^{- 2 U} (\mathscr{E} - \mathscr{E}_{\rm pot})^2 - m^2} \; \mbox{\boldmath$n$} =  \left(\int_{\rm mirror 1}^{\rm analyzer} -  \int_{\rm beam\; splitter}^{\rm mirror 2}\right) \sqrt{e^{- 2 U} (\mathscr{E} - \mathscr{E}_{\rm pot})^2 - m^2} \; \mbox{\boldmath$n$} \nonumber\\
& \approx  \left.l \mbox{\boldmath$h$} \cdot \mbox{\boldmath$\nabla$} \sqrt{e^{- 2 U} (\mathscr{E} - \mathscr{E}_{\rm pot})^2 - m^2}\right|_{\rm beam\; splitter}  \nonumber\\
& =  - \frac{\mathscr{E} - \mathscr{E}_{\rm pot,0}}{\sqrt{(\mathscr{E} - \mathscr{E}_{\rm pot,0})^2 e^{-2 U_0} - m^2}}  e^{-2 U_0} \mbox{\boldmath$h$} \cdot \mbox{\boldmath$\nabla$} \left((\mathscr{E} - \mathscr{E}_{\rm pot,0}) U +  \mathscr{E}_{\rm pot}\right) \, , 
\end{align}
where $\mbox{\boldmath$h$}$ is the vector connecting the beam splitter and the 1st mirror, $l$ is the length of the interferometer and $\mbox{\boldmath$\nabla$} U$ is evaluated at the position of the beam splitter (see Fig.~\ref{Fig:Interferometer}). The momentum of the particle at the beam splitter is $p_0 = \sqrt{(\mathscr{E} - \mathscr{E}_{\rm pot,0})^2 e^{-2 U_0} - m^2}$. We can choose the gravitational potential and the potential energy at the position of the beam splitter to vanish, $U_0 = 0$ and $\mathscr{E}_{\rm pot,0} = 0$. Then
\begin{equation}
\oint \sqrt{e^{- 2 U} (\mathscr{E} - \mathscr{E}_{\rm pot})^2 - m^2} \; \mbox{\boldmath$n$} = - l \frac{\mathscr{E}}{p_0} \mbox{\boldmath$h$} \cdot \mbox{\boldmath$\nabla$} \left(\mathscr{E} U + \mathscr{E}_{\rm pot}\right)  \, ,
\end{equation}
Here and in the following we take the length $l$ connecting the beam splitter and the 2nd mirror to be horizontal in the given coordinate system, that is, the beam splitter and mirror 2 have the same $r$--coordinate. When a particle is not charged $\mathscr{E}_{\rm pot} = 0$.

Non--relativistically, $\mathscr{E} \approx m$ and $p_0 \approx m v_0$, where $v_0$ is the group velocity of the matter wave at the beam splitter. Then
\begin{equation}
\oint \sqrt{e^{- 2 U} (\mathscr{E} - \mathscr{E}_{\rm pot})^2 - m^2} \; \mbox{\boldmath$n$} = - \frac{l}{v_0} \mbox{\boldmath$h$} \cdot \mbox{\boldmath$\nabla$} (m U + \mathscr{E}_{\rm pot}) \, ,
\end{equation}
For vanishing $\mathscr{E}_{\rm pot}$ this is the result of \cite{OverhauserColella74} which has been confirmed by Colella, Overhauser, and Werner \cite{COW75} using neutron interferometry. The appearance of the particle mass does not imply that the Weak Equivalence Principle is violated in interferometry or in quantum mechanics. Indeed, it has been shown by using quantum notions and measured expressions only that this phase shift can be given a natural form where the mass disappears. See \cite{Laemmerzahl96} for a thorough discussion of this aspect.

The first integral in \eqref{phasegeneralFlux} is the generalized Sagnac effect. It should be noted that it depends on the energy of the particle (frequency of the corresponding wave) only through the factor $\mathscr{E}$. The integral itself is a purely geometric expression related to properties of the Killing congruence. For small interferometers we obtain
\begin{equation}
\mathscr{E} \int_\Sigma e^{-U} \mbox{\boldmath$\omega$} \approx \mathscr{E} e^{- U_0} \omega_{\mu\nu} \Sigma^{\mu\nu} = \mathscr{E} \mbox{\boldmath$\omega$} \cdot \mbox{\boldmath$\Sigma$} \, ,
\end{equation}
where $\omega^\mu = \frac{1}{2} \epsilon^{\mu\nu\rho\sigma} u_\nu \omega_{\rho\sigma}$ and $\Sigma_\mu = \frac{1}{2} \epsilon_{\mu\nu\rho\sigma} u^\nu \Sigma^{\rho\sigma}$ ($\epsilon^{\mu\nu\rho\sigma}$ is the totally antisymmetric Levi--Civita symbol) are the rotation vector and area vector in $\mathscr{N}$, respectively. 

Since we consider the case that the potential varies very slowly over the interferometer the potential energy can be put in front of the integral
\begin{equation}
\int \mathscr{E}_{\rm pot} e^{- U} \mbox{\boldmath$\omega$} \approx \mathscr{E}_{\rm pot,0} \int e^{- U} \mbox{\boldmath$\omega$}  \, ,
\end{equation} 
so that for the case $\mathscr{E}_{\rm pot,0} = 0$ this term vanishes. (If we do not choose this condition, this term adds up with the first Sagnac term.)

With these results we obtain as total phase shift
\begin{equation}
\label{phaseshiftgen}
\Delta\Phi = \mathscr{E} \left(\mbox{\boldmath$\omega$} \cdot \mbox{\boldmath$\Sigma$} - \frac{l}{p_0} \mbox{\boldmath$h$} \cdot \mbox{\boldmath$\nabla$} \left(\mathscr{E} U + \mathscr{E}_{\rm pot}\right)\right) + e_p \mbox{\boldmath$B$} \cdot \mbox{\boldmath$\Sigma$} - g_p \mbox{\boldmath$E$} \cdot \mbox{\boldmath$\Sigma$} \ ,
\end{equation}
where $\mbox{\boldmath$E$}$ and $\mbox{\boldmath$B$}$ are the electric and magnetic field strength given by ${\mbox{\boldmath$B$}}_i = {\mbox{\boldmath$F$}}_{jk}$ and ${\mbox{\boldmath$E$}}_i = i {\check{\mbox{\boldmath$F$}}}_{jk}$ ($i, j, k$ cyclic).  
This is the general result for a small interferometer. Now we have to insert the particular gravitational field given by the Pleba\'{n}ski--Demia\'{n}ski metric and we also have to characterize the orientation of the interferometer. Since the baseline of the interferometer is always horizontal, there are only two degrees of freedom for the orientation of $\mbox{\boldmath$\Sigma$}$. These two angles are the angle of the baseline with respect to the line $\theta = const$, and the tilt of the interferometer. These two degrees of freedom then have to be matched to the longitudinal position of the laboratory on the surface of the Earth.

Let $\gamma$ be the angle of the baseline with respect to ${\mbox{\boldmath$e$}}_{\hat\varphi}$ and $\beta$ the tilt angle. Then, in the local normalized 3--bein position of the laboratory
\begin{eqnarray}
\mbox{\boldmath$h$} & = & h \cos\beta {\mbox{\boldmath$e$}}_{\hat r} - h \cos\gamma \sin\beta {\mbox{\boldmath$e$}}_{\hat \theta} - h \sin\gamma \sin\beta {\mbox{\boldmath$e$}}_{\hat \varphi} \\
\mbox{\boldmath$\Sigma$} & = & \Sigma \sin\beta {\mbox{\boldmath$e$}}_{\hat r} + \Sigma \cos\gamma \cos\beta {\mbox{\boldmath$e$}}_{\hat \theta} + \Sigma \sin\gamma \cos\beta {\mbox{\boldmath$e$}}_{\hat \varphi} \, .
\end{eqnarray}
With this specification the phase shift \eqref{phaseshiftgen} reads
\begin{eqnarray}
\label{phaseshiftnonrot}
\Delta\Phi & = & \mathscr{E} \Sigma \Biggl[- \frac{\mathscr{E}}{p_0} \left(\cos\beta a_{\hat{r}} - \cos\gamma \sin\beta a_{\hat{\theta}} - \sin\gamma \sin\beta a_{\hat{\varphi}}\right) \nonumber\\
& & \qquad\qquad - \frac{1}{p_0} \left(\cos\beta \partial_{\hat{r}} \mathscr{E}_{\rm pot} - \cos\gamma \sin\beta \partial_{\hat{\theta}} \mathscr{E}_{\rm pot} - \sin\gamma \sin\beta \partial_{\hat{\varphi}} \mathscr{E}_{\rm pot}\right) \nonumber \\
& & \qquad \qquad + \sin\beta \omega_{\hat{\theta}\hat{\varphi}} +  \cos\gamma \cos\beta \omega_{\hat{\varphi}\hat{r}} + \sin\gamma \cos\beta \omega_{\hat{r}\hat{\theta}} \Biggr] \nonumber \\
& & \qquad\qquad + e_p \Sigma (\sin\beta B_{\hat{r}} + \cos\gamma \cos\beta B_{\hat{\theta}} + \sin\gamma \cos\beta B_{\hat{\varphi}}) \nonumber \\
& & \qquad\qquad - g_p \Sigma (\sin\beta E_{\hat{r}} + \cos\gamma \cos\beta E_{\hat{\theta}} + \sin\gamma \cos\beta E_{\hat{\varphi}}) \ ,
\end{eqnarray}
where $\partial_{\hat{\mu}} = e^\nu_{\hat\mu} \partial_\nu$. This is the general result valid for a large class of interferometer orientations. 

From the fact that we can independently vary the angles $\beta$ and $\gamma$, we can extract from phase shift measurements the following combinations of terms
\begin{align}
\label{betagamma1}
\Delta\Phi(\beta = 0, \gamma = 0) & =  \mathscr{E} \Sigma \left[\omega_{\hat{\varphi}\hat{r}} - \frac{\mathscr{E}}{p_0} a_{\hat{r}}\right] - \frac{\mathscr{E} \Sigma }{p_0} \partial_{\hat{r}} \mathscr{E}_{\rm pot} + e_p \Sigma B_{\hat{\theta}} - g_p \Sigma E_{\hat{\theta}} \ , \\
\label{betagamma2}
\Delta\Phi(\beta = \tfrac{\pi}{2}, \gamma = 0) & = \mathscr{E} \Sigma \left[\omega_{\hat{\theta}\hat{\varphi}} + \frac{\mathscr{E}}{p_0} a_{\hat{\theta}}\right] + \frac{\mathscr{E} \Sigma }{p_0} \partial_{\hat{\theta}} \mathscr{E}_{\rm pot} + e_p \Sigma B_{\hat{r}} - g_p \Sigma E_{\hat{r}} \ , \\
\label{betagamma3}
\Delta\Phi(\beta = 0, \gamma = \tfrac{\pi}{2}) & = \mathscr{E} \Sigma \left[\omega_{\hat{r}\hat{\theta}} - \frac{\mathscr{E}}{p_0} a_{\hat{r}}\right] - \frac{\mathscr{E} \Sigma }{p_0} \partial_{\hat{r}} \mathscr{E}_{\rm pot} + e_p \Sigma B_{\hat{\varphi}} - g_p \Sigma E_{\hat{\varphi}} \ , \\ 
\label{betagamma4}
\Delta\Phi(\beta = \tfrac{\pi}{2}, \gamma = \tfrac{\pi}{2}) & = \mathscr{E} \Sigma \left[\omega_{\hat{\theta}\hat{\varphi}} + \frac{\mathscr{E}}{p_0} a_{\hat{\varphi}}\right] + \frac{\mathscr{E} \Sigma }{p_0} \partial_{\hat{\varphi}} \mathscr{E}_{\rm pot} + e_p \Sigma B_{\hat{r}} - g_p \Sigma E_{\hat{r}} \ .
\end{align}
Here the terms in the square brackets are gravito--inertial effects while the other terms are related to the electric and magnetic charges of the particles. In both cases we have terms of similar structure: The first terms, e.g., $a_{\hat{r}}$ and $\partial_{\hat{r}} \mathscr{E}_{\rm pot}$ are related to the change of the kinetic energy of the particle due to the interaction with the gravitational and electromagnetic field, while the other terms are Aharonov--Bohm type terms describing the coupling of the particle with the rotation flux or the magnetic and electric fluxes of the electromagnetic field. Note that each term is manifestly gauge invariant. 

It is obvious that charged particles have direct access to the electromagnetic field: an electrically charged particle has access to the magnetic field via the Aharonov--Bohm type terms and to the electric field via the derivative of the potential energy where in the latter the Killing vector has to be used. Both contributions can be separately identified through a variation of the momentum of the particles. The same holds, {\it mutatis mutandis}, for magnetically charged particles. For charged as well as neutral particles the contributions from the gravito--inertial field are of the same structure. In order to separate gravito--inertial and electromagnetic contributions one has to use two types of particles with different charge--to--mass ratios. 

Now we can use Eq.~\eqref{phaseshiftnonrot} to explicitly calculate the phase shift for various observers in a Pleba\'{n}ski--Demia\'{n}ski space--time. We start with an observer which is naturally aligned to the coordinate system used for expressing the Pleba\'{n}ski--Demia\'{n}ski metric \eqref{metric_non_accelerating}. As a second example we take an observer who rotates in the given coordinate system. This can be accomplished by using a different Killing congruence.

\subsection{Adapted Killing observer}

The natural Killing observer is characterized by the Killing field $\xi^\mu = \delta^\mu_0$ in the coordinate system of \eqref{metric_non_accelerating}. From the previous subsection it is clear that with varying charges and velocities we have access to all components of $a_{\hat i}$, $\omega_{\hat i \hat j}$, $\partial_{\hat i} \mathscr{E}_{\rm pot}$, $E_{\hat i}$, and $B_{\hat i}$ separately. These quantities have been calculated in \eqref{acceleration1}-\eqref{rotation3},\eqref{B_Field},\eqref{E_Field}. By further varying the latitude $\theta$ and the height $r$ we then have independent access to the various parts of these quantities:  
\begin{equation} \label{DeterminedParameters}
\begin{split}
& m, \quad \Lambda, \quad \Lambda m, \quad q a, \quad 2 q^2 - e^2 - g^2, \quad a \frac{\alpha}{w}, \quad a q, \quad a \Lambda, \quad a m, \quad q, \quad \Lambda q, \\ 
& m q, \quad q a \frac{\alpha}{w}, \quad g, \quad q g \frac{\alpha}{w}, \quad q e, \quad a g \frac{\alpha}{w}, \quad a e, \quad e a  q \frac{\alpha}{w}, \quad e, \quad q e \frac{\alpha}{w}, \quad q g \ ,
\\ & a e \frac{\alpha}{w}, \quad a g, \quad g a q \frac{\alpha}{w} \ .
\end{split}
\end{equation}
It should be noted that owing to the fact that the phase shift is a scalar calculated from a loop integral, it is independent of the chosen coordinate system. By varying the orientation and position of the interferometer and employing the expansion of the metric it is thus possible to extract all these parameter combinations. This particular procedure does not imply any interpretation of these parameters. In various physical settings other parameters and parameter combinations will appear~\footnote{The same happens in, e.g., the discussion of effects within the PPN formalism: While for a Schwarzschild metric all effects are related to the mass (and this mass can also be given a purely geometrical meaning in terms of the surface of the horizon or a certain length--to--radius ratio) within the PPN formalism each physical effect like light bending, perihelion shift, red shift, etc, is characterized by a different combination of PPN parameters.}. The fact that parameter combinations appear has been discussed to some extent in \cite{GriffithsPodolsky06,GriffithsKrtousPodolsky06} in terms of deficit angles characterizing the conical singularities. Other parameter combinations appear if one calculates geometrically invariant parameters \cite{GriffithsKrtousPodolsky06} which, in general, do not have to coincide with the combination of parameters leading to physical effects like light deflection, perihelion shift, Lense--Thirring effect, etc. Our particular setup and the possibility to vary the experimental setting leads to the set of measurable parameter combinations \eqref{DeterminedParameters}. 

Combinations of these expressions give $m$, $\Lambda$, $a$, $q$, $e$, $g$, and $\alpha/w$ separately. If the experiment is carried out with {\it neutral} particles, we have access to $a_{\hat i}$ and  $\omega_{\hat i \hat j}$ only and, thus, to the quantities in the first line of \eqref{DeterminedParameters} from which we can explicitly determine the parameter combinations
\begin{equation}
m \, , \quad \Lambda \, , \quad q \, , \quad a \, , \quad g^2 + e^2 \ ,  \quad \frac{\alpha}{w} \ . \label{measurement3}
\end{equation}
For neutral particles there is no separate access to $e$ or $g$. Also the acceleration $\alpha$ appears only through the combination with the twist parameter.   

The accuracy of the measurements of the acceleration is $\Delta a_{\hat r}/a_{\hat r} \approx 10^{-9}$ \cite{PetersChungChu99} which on Earth implies an absolute accuracy of $\Delta a \approx 10^{-8}\;{\rm m/s}^2$, and for rotation there is an absolute accuracy of $\delta\omega \approx 10^{-9}\;{\rm s}^{-1}$ \cite{GustavsonLandraginKasevich00}. In the near future the height on Earth and even to satellites orbiting the Earth at, say, 10000 km height can be defined with a precision of within 1 cm using Satellite Laser Ranging SLR \cite{Degnan07}. This gives the relative error of distance estimations $\Delta r/r \approx 10^{-9}$. 

All experimental results so far are in accordance with Newtonian gravity and Galilean kinematics both encoded in the mass parameter $m$ and the rotation $\mbox{\boldmath$\omega$}$. That implies that the mass and the rotation of the Earth can be determined with an accuracy of $10^{-8}$ and $10^{-4}$, respectively. The accuracy is also good enough to detect gravitationally the presence of a mass of 100 kg located at a distance of $\sim 1\;{\rm m}$ to the interferometer. 

Since all measurements are compatible with the Schwarzschild part of the Pleba\'{n}ski--Demia\'{n}ski metric we can place estimates, that is, maximum values, for the various Pleba\'{n}ski--Demia\'{n}ski parameters. Performing interference experiments on Earth as well as on a satellite 10000 km above the Earth surface~\footnote{The gravitational field on Earth contains of course smaller additional contributions from the Sun and the Moon, for example. As a consequence one should consider the combined gravitational field leading to small modifications of the gravitational field considered here. However, since all the Pleba\'{n}ski--Demia\'{n}ski parameters except the mass are very small, the corresponding modifications of the gravitational field related to the parameters $\Lambda$, $q$, $a$, $e$, and $g$ are safely negligible within our approximative scheme employed at this stage. Only modifications of the ``Newtonian'' part related to $m$ have to be taken into account. These modifications can be calculated easily and are given, e.g., by ephemerides. It is no problem to include these modifications in our discussion of experiments. However, since we are more interested in the principle question of accessibility to the Pleba\'{n}ski--Demia\'{n}ski parameters and in order not to blow up the formulas, we discuss the simpler question of measuring the gravitational field of the Earth only, neglecting all other masses in the Solar system.} and comparing the results would give an estimate $\Lambda \leq 10^{-32}\;{\rm m}^{-2}$ (this is four orders of magnitude better than what one obtains from redshift experiments \cite{KagramanovaKunzLaemmerzahl06}). If we take the NUT parameter $q$ to be of dimension kg (which might be justified since it has the interpretation of a gravitomagnetic mass), then we obtain the estimate $q \leq 10^{24}\;{\rm kg}$. From an asymptotic expansion of the metric and the requirement
that the $g_{0i}$ term should be of order $c^{-3}$ we infer that $a$ has the dimension ${\rm m}^2/{\rm s}$. This leads to an estimate $a \leq 10^{14}\;{\rm m}^2/{\rm s}^{-1}$. According to the previous settings the acceleration parameter $\alpha/w$ has the dimension ${\rm m}^{-1}{\rm s}^{-1}$ and, thus, we obtain the estimate $\alpha/w \leq 10^{-18}\;{\rm m}^{-1}{\rm s}^{-1}$. 


\subsection{Actively rotating observer}

Until now the interferometer is assumed to be positioned at a fixed point given by the coordinates $(r, \theta, \varphi)$. In realistic situations the laboratory and, thus, the interferometer is attached to the surface of the Earth. Therefore both are co--rotating with the Earth. That introduces an additional angular velocity. This can be described as follows: Beside the time--like Killing vector $\xi$ there is also a Killing vector related to the axial symmetry. In adapted coordinates this Killing vector is given by $\eta^\mu = \delta^\mu_\varphi$. The sum of two Killing vectors again is a Killing vector. Therefore, a stationary rotating laboratory on the surface of the Earth can be described by the Killing vector $\xi^\prime = \xi + \omega_\oplus \eta$, where $\omega_\oplus$ is the angular velocity of the rotating Earth. This new Killing vector now gives a modified total rotation
\begin{align}
\omega_{{\hat r} {\hat\theta}}^\prime & =  0 \\
\omega_{\hat\theta\hat\varphi}^\prime & \approx  \omega_{\hat\theta\hat\varphi} - \frac{1}{2}\omega_\oplus \cos\theta \Bigg( 2\left(1+\frac{m}{r}\right) + \Lambda r\left(m + \frac{r}{3}\right)  
+ 2\frac{\alpha}{w}r(2q + a\cos\theta) \Bigg)
 \\
\omega_{{\hat r} \hat\varphi}^\prime & \approx  \omega_{{\hat r} \hat\varphi} - \omega_\oplus \sin\theta \left(1 + \frac{1}{2} r \Lambda (2 m + r) + r \frac{\alpha}{w}(3q + a\cos\theta)\right) \\
\omega_{{\hat r} {\hat t}}^\prime & \approx  a\omega_\oplus \sin^2\theta \left( \frac{m}{r^2} - \frac{\Lambda r}{3} \right) \\
\omega_{\hat\theta {\hat t}}^\prime & \approx  \omega_\oplus \sin\theta \left(\frac{q}{r} - \left(\frac{r\Lambda}{6} + \frac{m}{r^2}\right)(q + 2 a \cos\theta) - q\frac{a}{w}\alpha\cos\theta \right)  \ .
\end{align}
as well as a modified acceleration
\begin{align}
a_{\hat r}^\prime & \approx  a_{\hat r} - 2\omega_\oplus a\sin^2\theta \left( \frac{m}{r^2}-\frac{r\Lambda}{3} \right) 
\nonumber \\
&  \qquad - \omega_\oplus^2 \left(r+m +\frac{2}{3}r^2\Lambda(3m+r) + 3\frac{q}{w}r^2\alpha \right) \sin^2\theta \\
a_{\hat\theta}^\prime & \approx  a_{\hat\theta} + \omega_\oplus a \sin2\theta \left(\frac{2m}{r^2} + \frac{r\Lambda}{3} \right)- 2 \frac{q}{r}
\omega_\oplus\sin\theta 
\nonumber \\
&  \qquad -
\omega_\oplus^2\sin\theta\cos\theta \left( r+2m + \frac{r^2\Lambda}{3}(r + 4m) 
+ 2 \frac{q}{w}\alpha r^2 \right)   \\
a_{\hat\varphi}^\prime & =  0 \ .
\end{align}
One easily recognizes the standard purely rotating observer in $\omega_{\hat\theta\hat\varphi}^\prime$ and $\omega_{\hat r\hat\varphi}^\prime$ if one sets all parameters in the Pleba\'{n}ski--Demia\'{n}ski solution to zero.  

This has to be inserted into the general result \eqref{phaseshiftnonrot}. As far as the access to the Pleba\'{n}ski--Demia\'{n}ski parameters is concerned, we obtain the same result so that we omit to present the lengthy formulas.

\section{Conclusion}

We calculated the phase shift for a charged particle interference experiment in a general Pleba\'{n}ski--Demia\'{n}ski black hole space--time. In doing so we put emphasis on a gauge invariant implementation of the symmetry conditions which are needed in order to obtain a stationary interference pattern. The gauge invariance ensures that each term in the phase shift is gauge invariant and, thus, has a clear physical interpretation. 

Besides addressing the issue of an appropriate formalism to describe such experiments we also answered the question whether it is possible to have access to all parameters characterizing the family of Pleba\'{n}ski--Demia\'{n}ski generalized black hole solutions. The result \eqref{measurement3} shows that the phase shift for a Mach--Zehnder interferometer in such Pleba\'{n}ski--Demia\'{n}ski space-times is influenced by all parameters characterizing a general accelerating black hole defined by the metric \eqref{metric_non_accelerating}. By varying the orientation, the latitude and the height in the given gravitational field in principle it is possible to have access to all parameters of the Pleba\'{n}ski--Demia\'{n}ski solution. 

The above results may also be used to describe atom interferometric experiments. In the non--relativistic limit the phase shift in a Kasevich--Chu like interferometer \cite{KasevichChu92,PetersChungChu99} is given by $\Delta\Phi = - \mbox{\boldmath$k$} \cdot \left(\mbox{\boldmath$a$} + \mbox{\boldmath$v$} \times \mbox{\boldmath$\omega$}\right) T^2$ where $\mbox{\boldmath$k$}$ is the wave vector of the lasers serving as beam splitter and $T$ is the time--of--flight of the atoms between two laser pulses. $\mbox{\boldmath$a$}$ and $\mbox{\boldmath$\omega$}$ are the acceleration and rotation of the interferometer which can be taken to be the corresponding quantities calculated above. 

\subsection*{Acknowledgement}

We like to thank H. Dittus and V. Perlick for enlightening discussions. V.K. thanks the German Academic Exchange Service DAAD and C.L. the German Aerospace Center DLR for financial support.  



\end{document}